\documentclass{aip-cp}

\usepackage[authoryear]{natbib}
\usepackage{rotating}
\usepackage{graphicx}

\newcounter{daggerfootnote}
\newcommand*{\daggerfootnote}[1]{%
    \setcounter{daggerfootnote}{\value{footnote}}%
    \renewcommand*{\thefootnote}{\fnsymbol{footnote}}%
    \footnotetext[2]{#1}%
    \setcounter{footnote}{\value{daggerfootnote}}%
    \renewcommand*{\thefootnote}{\arabic{footnote}}%
    }

\begin{document}

\title{A Long-Term Spectral Study of the Single Active Giant \newline OP Andromedae}

\author[aff1]{S. Georgiev\corref{cor1}}
\author[aff1]{R. Konstantinova-Antova}
\author[aff1]{A. Borisova$^{\dagger}$}
\author[aff1]{D. Kolev}
\author[aff2]{M. Auri\`{e}re}
\author[aff2]{P. Petit}
\author[aff1]{M. Belcheva}
\author[aff1]{H. Markov}
\author[aff1]{R. Bogdanovski}
\author[aff1]{B. Spassov}
\author[aff1]{R. Zamanov}
\author[aff1]{N. Tomov}
\author[aff1]{A. Kurtenkov}

\affil[aff1]{Institute of Astronomy and NAO, Bulgarian Academy of Sciences, 72 Tsarigradsko shose, 1784 Sofia, Bulgaria}
\affil[aff2]{Universit\'e de Toulouse, Institut de Recherche en Astrophysique et Plan\'etologie, 14 avenue \'Edouard Belin, 31400 Toulouse, France}
\corresp[cor1]{Corresponding author:sgeorgiev@astro.bas.bg}

\protect\daggerfootnote{Deceased on 11 December 2017}

\maketitle

\begin{abstract}
We present a spectral study of the single magnetically active K giant OP And in the period 1979 -- 2018, monitoring the variability of the activity indicator line H${\alpha}$. Original data obtained in the period 2015 -- 2018 with the echelle spectrograph \textit{ESpeRo} at the 2m telescope of the National Astronomical Observatory Rozhen in Bulgaria, previously unpublished original data obtained in the period 1997 -- 2007 and on one night in 2013 with the Coude spectrograph at the same telescope, as well as data from the literature are presented in this study. The variability of the H${\alpha}$ line reveals that the activity level of OP And is higher in the period 1993 -- 2000, while during the period 2008 -- 2010 it is lower, possibly close to a minimum. Also, our data for the period 2015 -- 2018 show that the activity level is increasing again. Spectral observations of the activity indicators CaII H\&K lines and CaII IR triplet are sparse during the studied period. We use such ones when possible to confirm the detection of some flare events. The structure of H${\alpha}$ changes with the activity level: when the activity is higher, we observe a blue-shifted component of this line, interpreted as an expanding area above the photosphere, but during a lower activity period it is almost absent. Our results are in a good agreement with the idea that the magnetic field controls the mass outflow in this giant. More years of observations are necessary to determine the eventual activity cycle of OP And.
\end{abstract}

\section{INTRODUCTION}
OP And (HD 9746) is a single active giant of spectral class K1III. Its apparent magnitude is 6.21$^m$ -- 6.47$^m$ \citep{samus2011}, and the color index $B - V = 1.21^m$ \citep{deMM1999}. The brightness of the star varies by  $\sim$0.1$^m$ in V filter of the Johnson-Cousins UBVRI filter system and these changes are of variable period \citep{sh1988}. This variability is thought to be due to photospheric spots which change their configuration and distribution in time. The rotational period is $\mathrm{P_{rot}}$ = 76 d \citep{sh1988, ka2005}. The mass of this star is $2.0 \pm 0.5$ M$_{\odot}$, the effective temperature is $\mathrm{T_{eff}}$ = 4420 K, the surface gravity log\textit{g} = 2.3 \citep{auriere2015} and the metallicity [Fe/H] $= 0.0$ \citep{bala2000}. This giant is known for its high lithium abundance: log${\epsilon}$(Li)= 3.5 \citep{drake2002}. For comparison, late giants with near-solar metallicity are expected to have log${\epsilon}$ $\leq$ 1.5 \citep{iben1967}: OP And is member of a rare group of lithium-rich giants. The projected rotational velocity is vsini = 8.7 $\pm$ 1.0 km/s, and the radial velocity RV = -42.7 $\pm$ 0.3 km/s \citep{deMM1999}. No variations are measured for this value, suggesting that OP And is a single star. The surface-averaged magnetic field is $\mathrm{B_{mean}}$ = 15 G \citep{auriere2015}. The correlation of the magnetic field strength and the rotational period suggest that the reason for the activity in OP And is a magnetic dynamo \citep{auriere2015}. The evolutionary stage of this star is at the end of the first dredge-up phase and the beginning of the red giant branch, at the end of the so-called first magnetic strip \citep{charbonnel2017}. Only one other giant, V1192 Ori, is at this evolutionary stage and is known to have fast rotation and a surface magnetic field \citep{auriere2015}. \newline The spectrum of OP And shows strong emission cores of variable intensity in the CaII H\&K lines (3968.5 and 3933.7 {\AA} respectively), which lie inside deep absorption \citep{strass1990}. The emission intensity is comparable to that of active binaries of type RS CVn. The CaII infrared triplet (CaII IRT) lines at 8498, 8542 and 8662 {\AA}, which are one of the activity indicators \citep{dempsey1993} also vary in intensity. The H${\alpha}$ spectral line has a complex profile in the spectrum of OP And: an emission core is present which fills the absorption component to a different extent; a 79 km/s blue-shifted emission component is also present, which is explained with a hot expanding area above the photospere \citep{mihalas1978} and is linked to mass loss. In some cases of flares, a red-shifted emission component is also observed for this line \citep{ka1995}. \newline The chromospheric flares of OP And have been detected by photometric \citep{kaa2000} and spectroscopic observations (in H${\alpha}$) \citep{ka2001, borisova2012}. A previous study of the long-term variability of this star \citep{borisova2012} showed that the activity level was higher in 1993 -- 1996 based on the spectral indicators, while in 2008 and 2010 it was lower. No long-term activity indicators period is known for this star so far.

\section{OBSERVATIONS AND DATA PROCESSING}
In our study we present observational spectral data of OP And in the period 1979 -- 2018. We use the H${\alpha}$ line as an activity indicator. Our observations in the spectral regions of CaII H\&K and CaII IRT are fewer and we use these lines as secondary indicators only to confirm possible flares of OP And. \newline We used original data obtained in the period September 2015 -- August 2018 with the echelle spectrograph ESpeRo at the 2m telescope of the Rozhen National Astronomical Observatory (NAO), Bulgaria \citep{bonev2017}, as well as previously unpublished data obtained with the Coude spectrograph at the same telescope \citep{kolev1996} in the period January 1997 -- November 2007 and on one night in August 2013. A log of our observations is presented in Table \ref{table:obs}. We also use results of previous spectral observations of OP And from the literature. Such observations in the region of H${\alpha}$ were first done in 1979 by \citet{fekel1986} with the 2.1m telescope of McDonald Observatory with spectral resolution R $\approx$ 15000. Following observations in the region of H${\alpha}$ were done in 1983 by \citet{strass1990} with the same telescope and spectral resolution. We note that the data from 1979 and 1983 only consists of single observations, so we cannot use them when trying to determine an eventual long-term variability period; however we include them in our study because these spectra confirm the profile of H${\alpha}$ typical for this star. In the period 1993 -- 1996 spectra of the H${\alpha}$ line were obtained by \citet{ka2001} with the Coude spectrograph at the 2m telescope of NAO with R $\approx$ 20000. Observations were done in 2008 and 2010 \citep{borisova2012} with the NARVAL spectropolarimeter at the 2m TBL at Pic du Midi Observatory, France with spectral resolution of 65000; detailed description of this instrument can be found in \citet{auriere2003}. Our observations in the period 1997 -- 2007 and the night in August 2013 have R $\approx$ 20000 and the ones in 2015 -- 2018 have R $\approx$ 32000. The signal-to-noise ratio (SNR) for all our observations in the region of H${\alpha}$ is above 100, the typical value being about 200.\newline For the spectra we obtained, data reduction was done using the software package \textit{Image Reduction and Analysis Facility (IRAF)}. In cases where multiple spectra were obtained in one night or in consequtive nights we used their median spectrum to remove any random noise that could occur. This was done after inspection of the individual spectra for absence of any short-term changes. We use the resulting data to study the activity level of OP And by measuring the relative intensity $\mathrm{R_c}$ regarding the continuum in the core of the H${\alpha}$ line and its blue-shifted emission wing. However, spectra of late giants contain many absorption lines, so that a true continuum level is not present. We instead measured $\mathrm{R_c}$ regarding to a pseudo-continuum that we determined in accordance with the procedure described in \citet{eaton1995}.\newline The relative intensity is an instrument dependent value as higher spectral resolution results in deeper absorption lines, hence in lower $\mathrm{R_c}$. This problem has been resolved by lowering the spectral resolution of the spectra obtained with NARVAL. We then measured the relative intensity of the selected lines in the newly constructed spectra using the software package \textit{IRAF}.

\begin{table}
	\begin{tabular}{llllllll}
	\hline
	Date       & JD        & Spectrograph & SNR & Date       & JD        & Spectrograph & SNR \\ \hline
	1997-01-23 & 2450472.2 & Coude     & 200 & 2006-12-01 & 2454071.5 & Coude     & 120 \\
	1997-01-24 & 2450473.2 & Coude     & 200 & 2007-07-01 & 2454283.6 & Coude     & 200 \\
	1997-01-25 & 2450474.2 & Coude     & 200 & 2007-07-30 & 2454311.5 & Coude     & 270 \\
	1997-08-14 & 2450675.4 & Coude     & 170 & 2007-08-02 & 2454314.6 & Coude     & 270 \\
	1997-08-15 & 2450676.5 & Coude     & 200 & 2007-08-03 & 2454316.5 & Coude     & 270 \\
	1997-10-16 & 2450738.3 & Coude     & 300 & 2007-11-27 & 2454432.3 & Coude     & 300 \\
	1998-09-30 & 2451087.4 & Coude     & 270 & 2013-08-17 & 2456522.5 & Coude     & 170 \\
	1998-10-28 & 2451115.4 & Coude     & 150 & 2015-09-02 & 2457268.0 & ESpeRo    & 110 \\
	1999-01-10 & 2451189.2 & Coude     & 140 & 2015-10-04 & 2457300.4 & ESpeRo    & 180 \\
	1999-03-03 & 2451241.2 & Coude     & 200 & 2015-12-28 & 2457385.3 & ESpeRo    & 180 \\
	2000-06-24 & 2451720.6 & Coude     & 200 & 2016-01-27 & 2457415.3 & ESpeRo    & 150 \\
	2000-06-25 & 2451721.5 & Coude     & 250 & 2016-07-26 & 2457596.0 & ESpeRo    & 200 \\
	2000-07-11 & 2451737.5 & Coude     & 200 & 2016-09-18 & 2457650.4 & ESpeRo    & 100 \\
	2000-08-15 & 2451772.5 & Coude     & 180 & 2016-10-10 & 2457671.5 & ESpeRo    & 100 \\
	2000-09-09 & 2451797.5 & Coude     & 150 & 2016-11-15 & 2457708.0 & ESpeRo    & 170 \\
	2000-10-22 & 2451840.4 & Coude     & 150 & 2016-12-09 & 2457732.4 & ESpeRo    & 150 \\
	2005-06-27 & 2453548.5 & Coude     & 120 & 2017-06-30 & 2457935.5 & ESpeRo    & 200 \\
	2005-11-10 & 2453685.2 & Coude     & 110 & 2018-01-06 & 2458125.3 & ESpeRo    & 130 \\
	2006-08-09 & 2453957.5 & Coude     & 180 & 2018-03-07 & 2458185.2 & ESpeRo    & 200 \\
	2006-08-12 & 2453960.4 & Coude     & 200 & 2018-08-24 & 2458355.5 & ESpeRo    & 150 \\
	2006-10-03 & 2454012.4 & Coude     & 200 & & & & \\ \hline
	\end{tabular}
	\caption{Table of observations}
	\label{table:obs}
\end{table}

\section{RESULTS AND DISCUSSION}
\begin{figure}[!htb]
	\begin{minipage}{0.7\textwidth}
		\includegraphics[scale=0.4, angle=270]{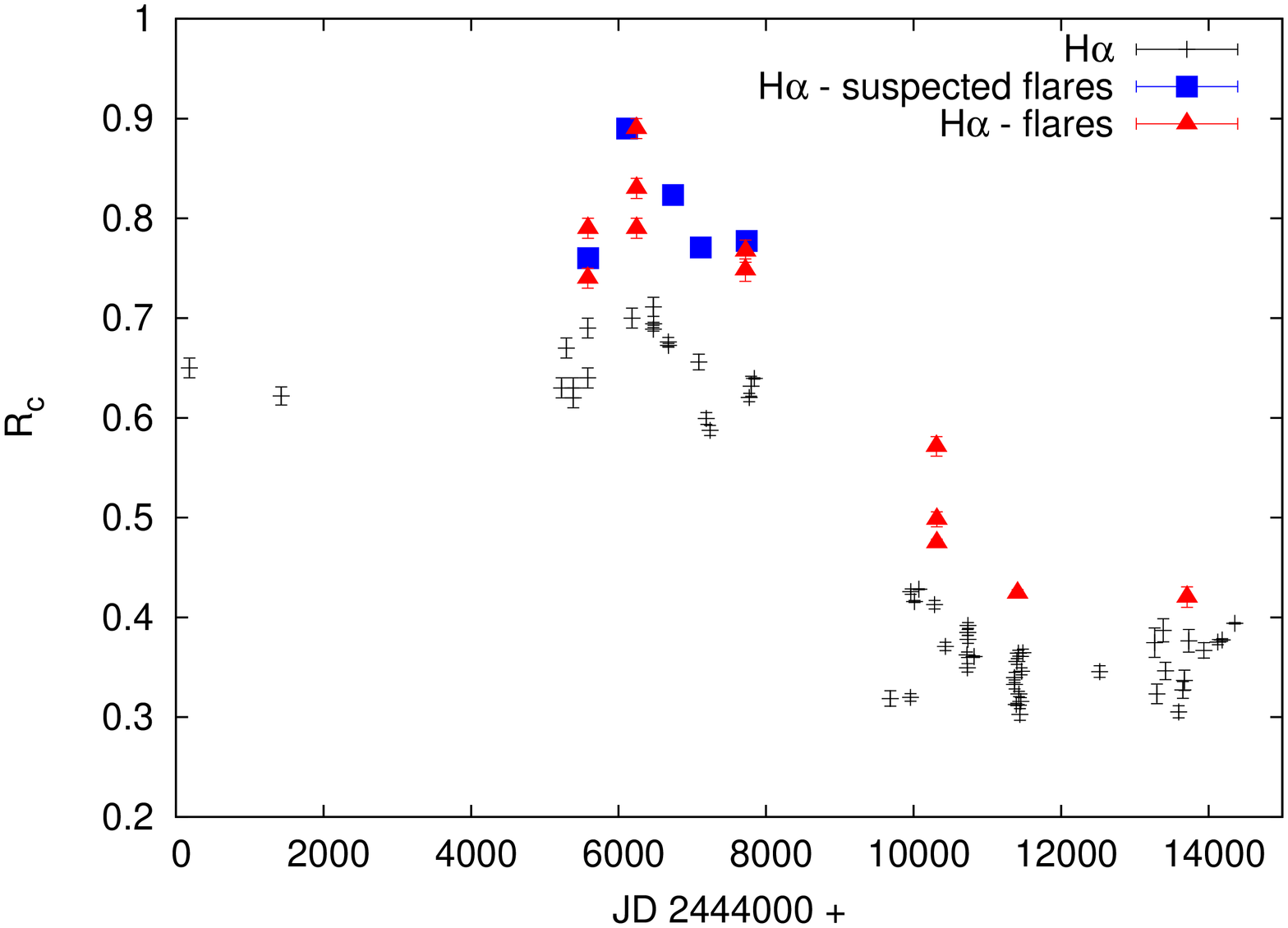}
		\includegraphics[scale=0.4, angle=270]{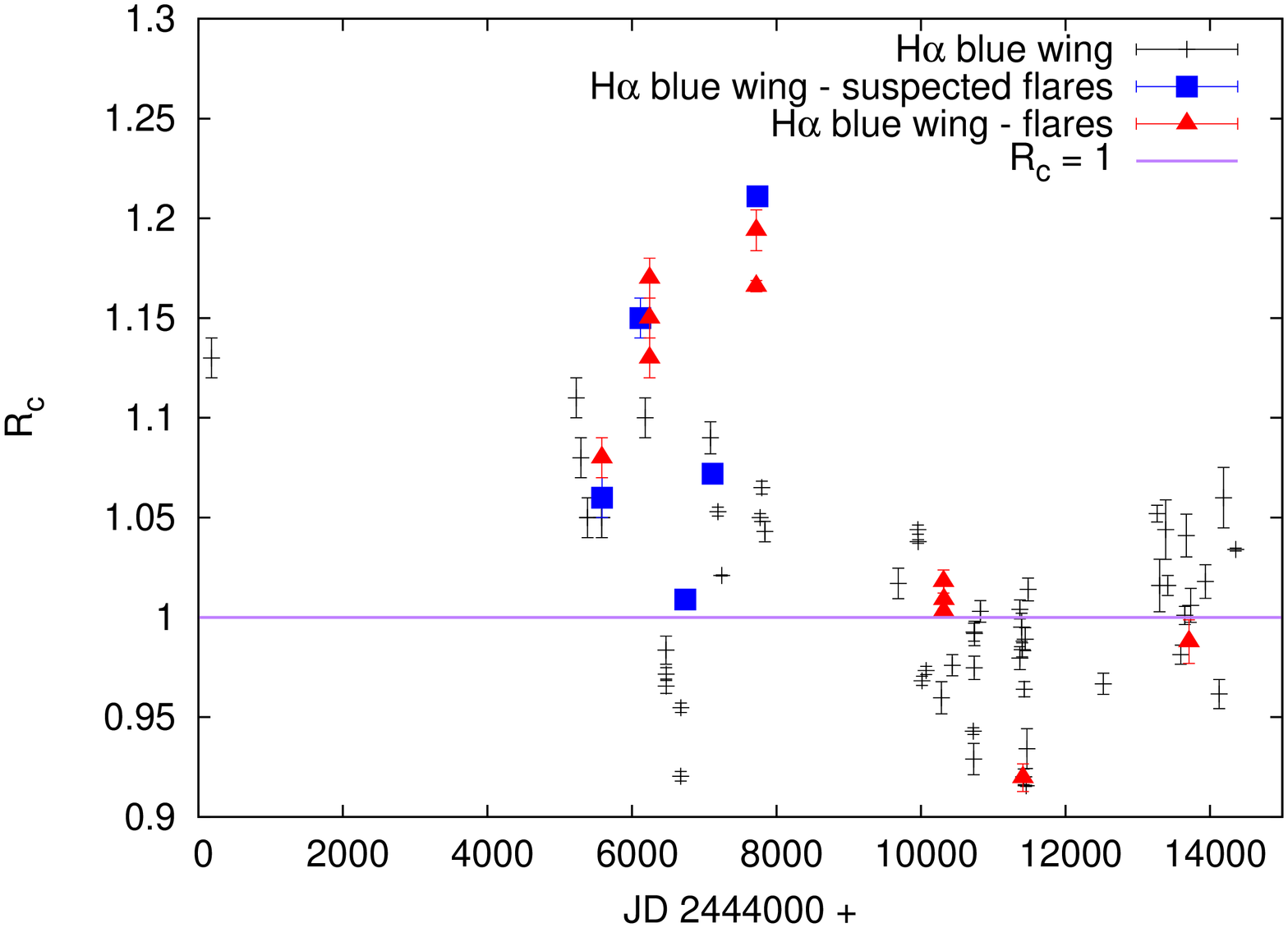}
	\end{minipage}
	\label{fig:ha-both}
	\caption{Measured values of $\mathrm{R_c}$ for the H${\alpha}$ core (top) and blue-shifted emission wing (bottom) for OP And. Flares and suspected flares are designated by red triangles and blue squares respectively.}
\end{figure}

\begin{figure}[!htb]
	\begin{minipage}{0.35\textwidth}
		\includegraphics[scale=0.24, angle=270]{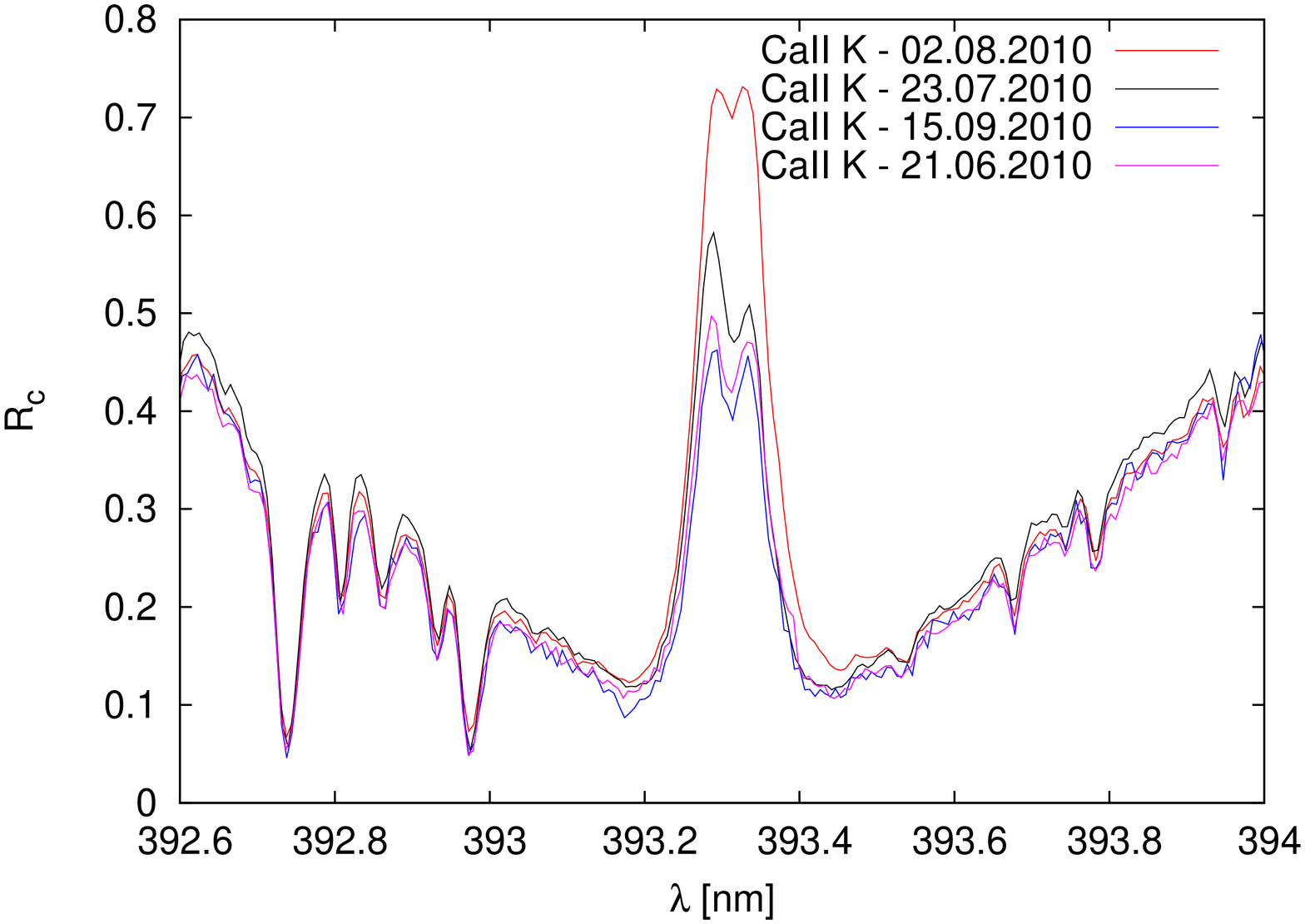}
		\includegraphics[scale=0.24, angle=270]{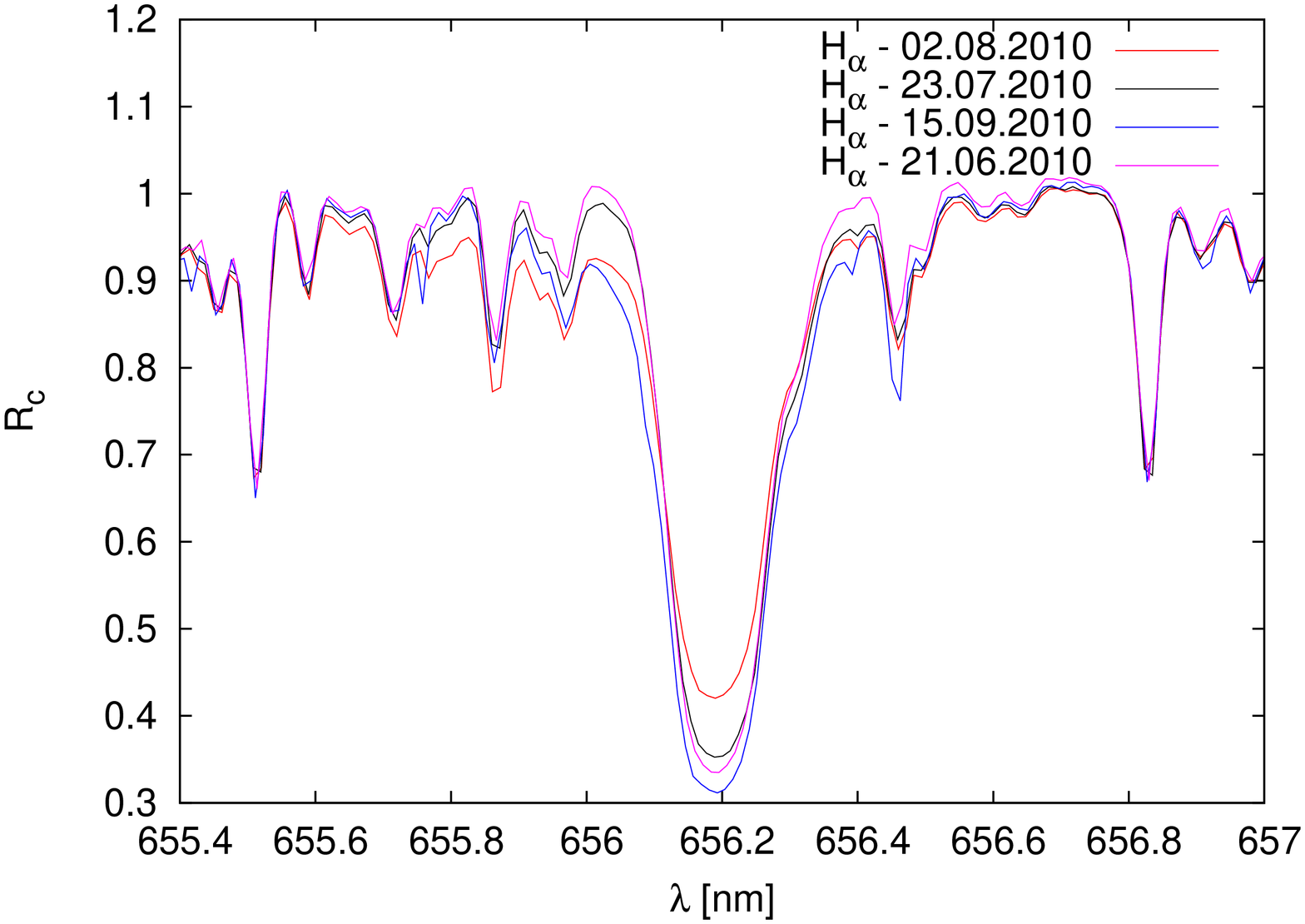}
	\end{minipage}
	\begin{minipage}{0.35\textwidth}
		\includegraphics[scale=0.24, angle=270]{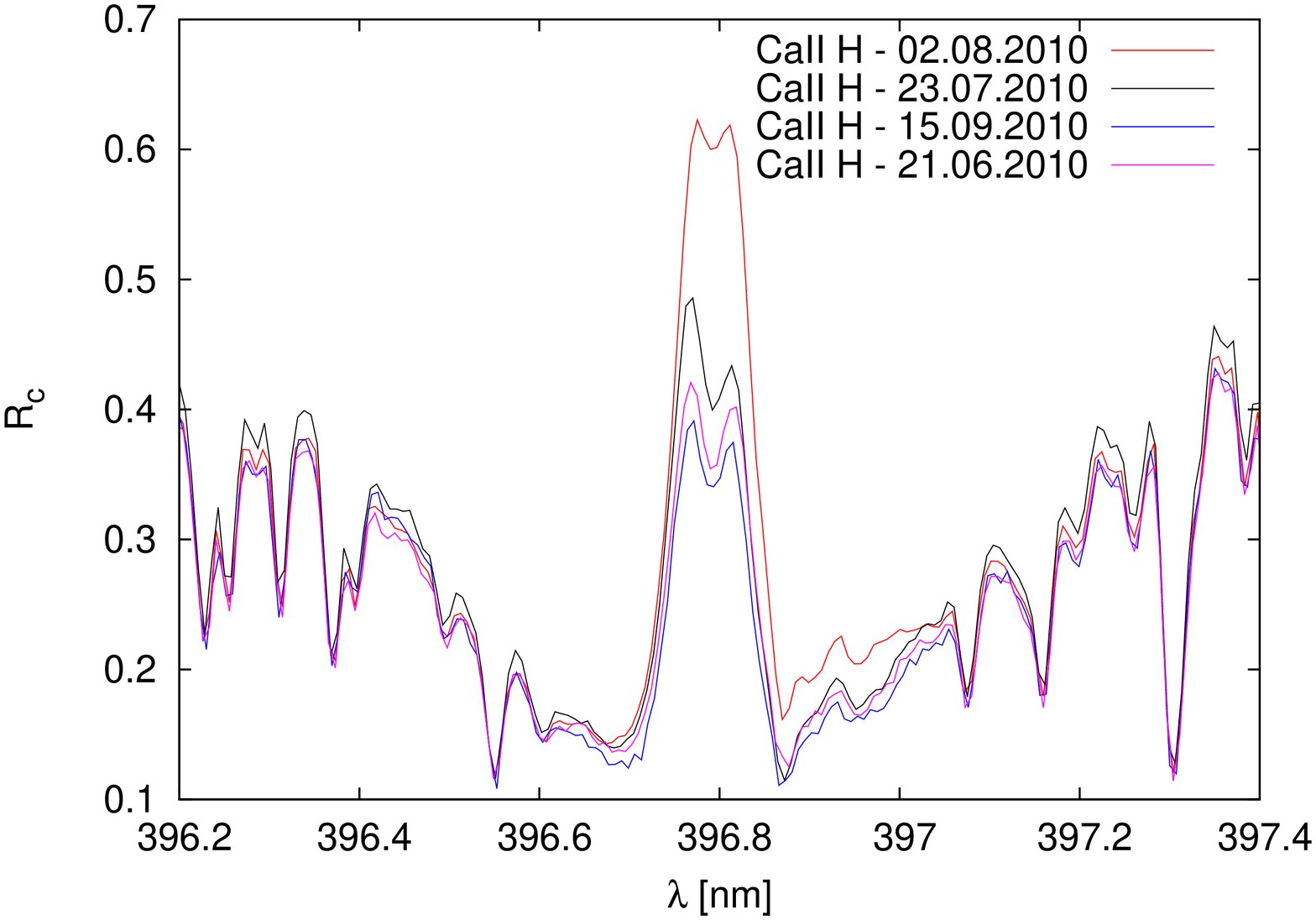}
		\includegraphics[scale=0.24, angle=270]{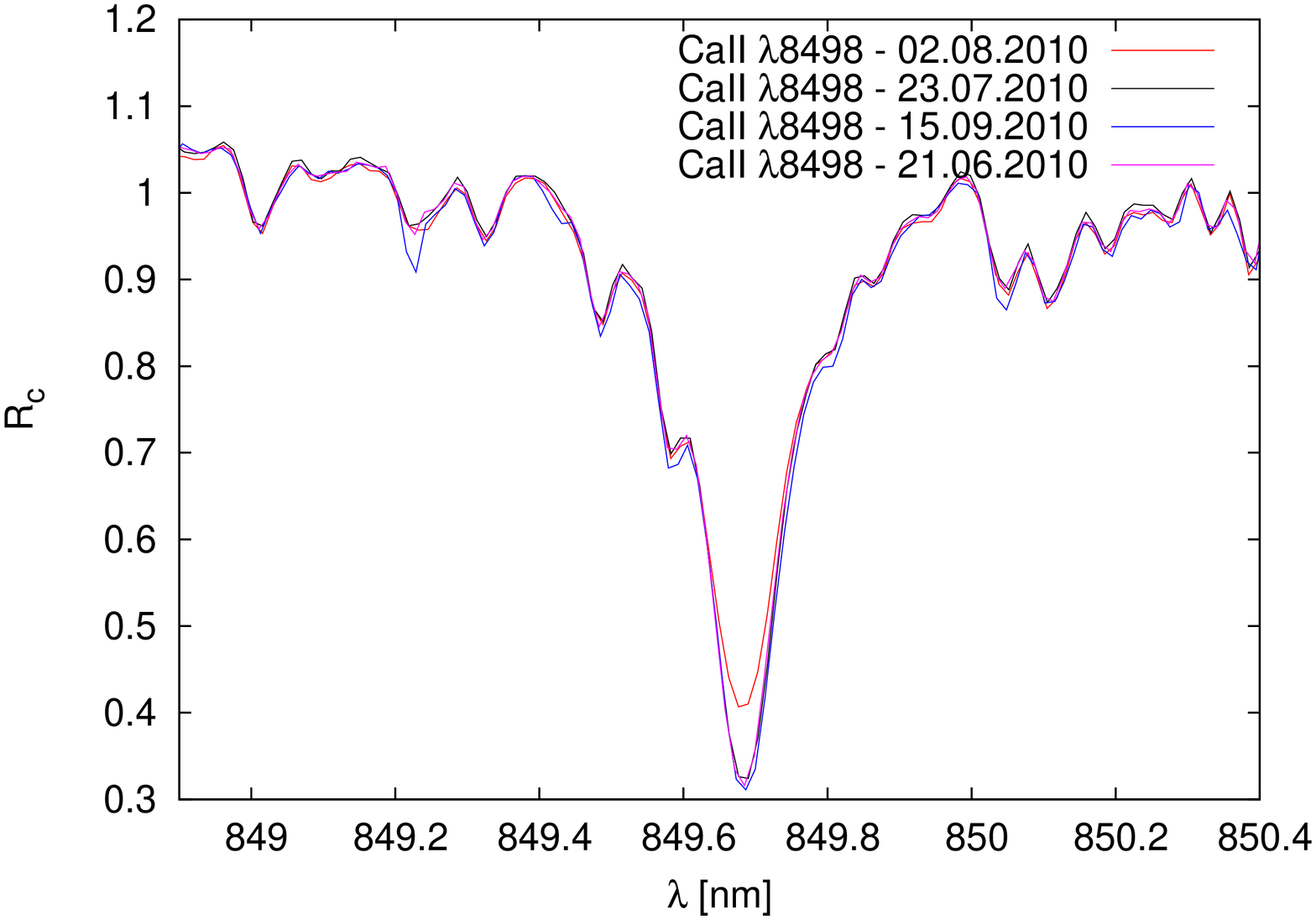}
	\end{minipage}
	\label{fig:flare}
	\caption{Spectrum of the flare on 2 August 2010 in the regions of four activity indicators (CaII H\&K lines, H${\alpha}$ and one line of the CaII IRT) is compared with other spectra of OP And where no flares are detected.}
\end{figure}

The measured values of $\mathrm{R_c}$ in the core and blue wing of H${\alpha}$ are presented in Figure \ref{fig:ha-both} in the top and bottom panel respectively. The suspected flare events are presented in blue squares. Only single observations in the H${\alpha}$ region are available for those events and we cannot be sure that they are flares. In the cases where we are sure that a flare has been registered, the measurements are presented in red triangles. We consider the detection of a flare a certainty in two types of cases: 1) when observations of other activity indicators together with H${\alpha}$ are available and show increased emission level (CaII H\&K and CaII IRT for the flare on 2 August 2010 and CaII IRT for the flare on 15 November 2016);  2) when several observations in the region of H${\alpha}$ were done consecutively and we observe variability typical for flare events (the flares on 18-19 August 1994, 8-9 August 1996, 24-25 June 2000 and 30 July -- 3 August 2007). As an illustration, the spectrum of OP And taken during the flare on 2 August 2010 in the regions of the selected activity indicators is presented in Figure \ref{fig:flare} and plotted over other spectra of this star from 2010 where no flares are detected. Changes in the relative intensity of the activity indicators are obvious during a flare event.\newline

\begin{figure}[!htb]
	\includegraphics[scale=0.3, angle=270]{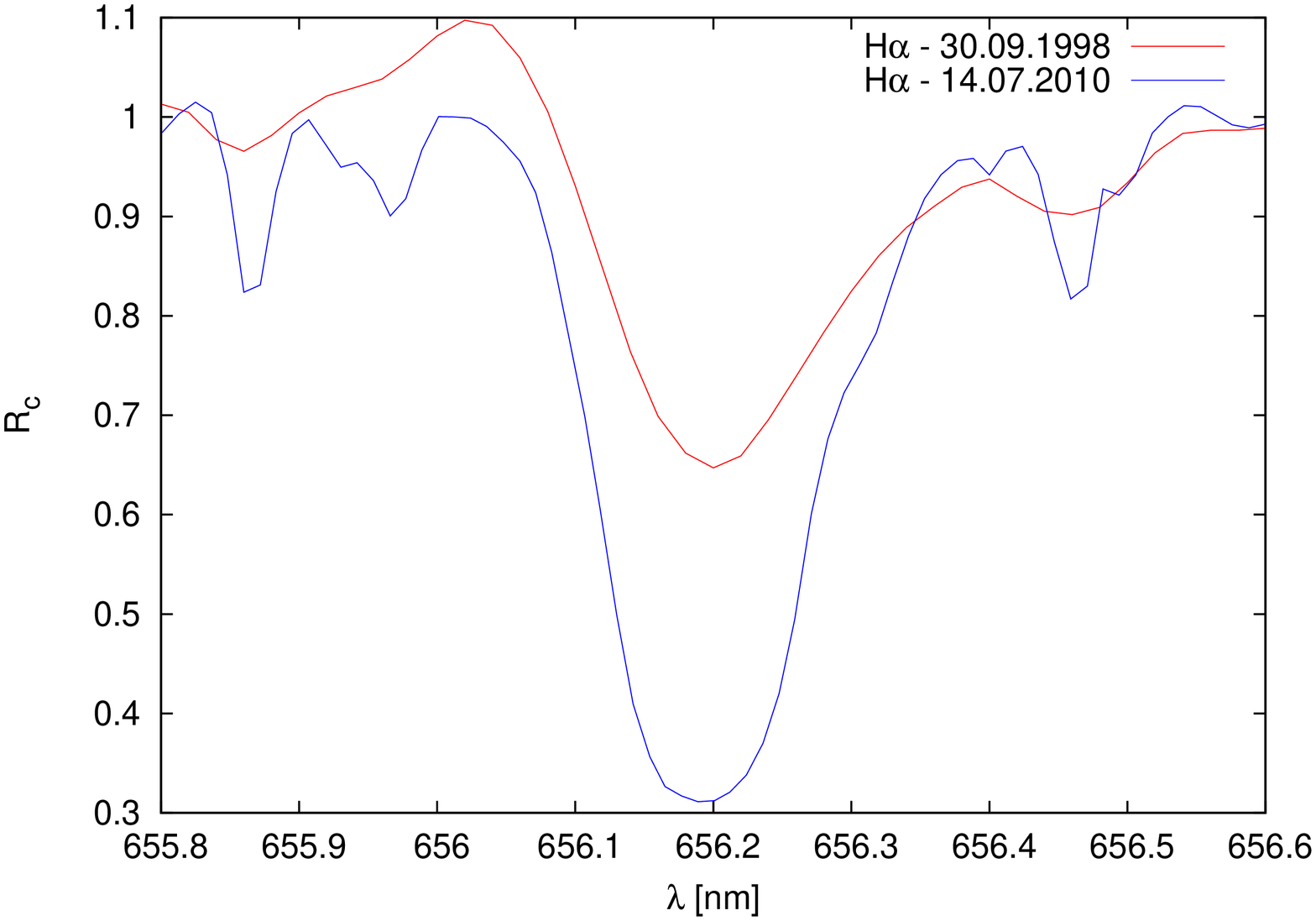}
	\label{fig:comp}
	\caption{Spectra of OP And obtained on 30 September 1998 (during a higher activity level) and on 14 July 2010 (during a lower activity level) are plotted together in the region of H${\alpha}$ for comparison. A clear difference is seen between the relative intensity at the core of the line; a strong blue-shifted emission wing is also apparent in the first spectrum, while in the second one it is indistinguishable from the continuum level.}
\end{figure}

Based on the presented values of $\mathrm{R_c}$ for both the H${\alpha}$ core and blue-shifted wing, we obtain the following results:
\begin{enumerate}
\item The long-term variability of OP And is confirmed. The observed activity level based on measurements of the H${\alpha}$ core is highest in the period of 1993 -- 2000, low in the years 2008 and 2010 and again higher in the period of 2015 -- 2018. As an illustration, a comparison between profiles of the H${\alpha}$ line is given in Figure \ref{fig:comp} where a spectrum taken on 30 September 1998 (during a higher activity level) is plotted over a spectrum taken on 14 July 2010 (during a lower activity level). No long-term period is apparent, however. It seems that such a period should be longer than the dataset available.
\item In the period of higher activity (1993 -- 2000) the blue-shifted emission component of the H${\alpha}$ line is present in most of the data. In the period of lowest observed activity (2008 and 2010) this component is only slightly above the continuum level or, in most cases, indistinguishable from it.
\item It seems, in the period of higher activity (1993 -- 2000) we detect more flare events than in the period of lower activity.
\end{enumerate}

As far as the blue-shifted emission component of H${\alpha}$ is linked to an expanding hot area in the atmosphere (which considering the low value of log\textit{g} leads to mass loss from this giant), we can suggest that the activity level possibly affects the mass loss: during the years of higher activity it is higher, while during the years of lower activity it is low or not present.\newline In the observed period of lower activity, two flare events were detected: on 30 July -- 3 August 2007 and on 2 August 2008. One more flare is observed on 15 November 2016. Even during the last three flares, no significant blue-shifted emission component is observed. In the period of higher activity three flares of OP And are detected: on 18-19 August 1994, 8-9 August 1996 and 24-25 June 2000; detection of five more flares is suspected on 23 August 1994, 2 February 1996, 16 October 1997, 28 October 1998 and 11 July 2000. The flares of this giant last long, i.e. for more than one day. This could be related to the height of the magnetic loops in the stellar atmosphere: see \citet{katsova1991}.

\section{CONCLUSION}
The long-term variability of the single active giant OP And in the period 1979 -- 2018 is studied. Variability in the H${\alpha}$ core and its blue-shifted emission wing is confirmed. In the period of higher activity 1993 -- 2000 this blue wing is distinctively above the continuum level and is relatively strong, while in the years of lowest observed activity 2008 and 2010 it is only slightly above the continuum level or not present. As this component is linked to mass outflow, it could be that the long-term activity level strongly affects the mass loss in this giant.\newline A total of six flares of OP And are detected and detection of five more such events is suspected. If these five events are indeed flares, it could be that during the period of higher activity OP And flares more often.\newline More observations are necessary in order to determine the eventual long-term activity period of this star.

\section{ACKNOWLEDGEMENTS}
The Narval observations in 2008 are granted under OPTICON program. The observations in 2010 are funded under Bulgarian NSF contract DSAB 02/3. S.G., R.K.A., M.B. and H.M. acknowledge partial financial support under Bulgarian NSF contract DN 08/1. H.M. and A.K. acknowledge partial financial support under Bulgarian NSF contract DN 18/13.



\end{document}